\documentclass[acus]{JAC2000}

\usepackage{graphicx}

\begin{document}
\title{\flushright{T15}\\[15pt] \centering
PROPOSAL FOR THE AEROGEL CHERENKOV
COUNTERS FOR THE PEP-N DETECTOR\thanks{Partially supported by Russian
Foundation for Basic Research, grants 99-02-16385, 99-02-16712.}}

\author{
M.Yu.Barnyakov, V.S.Bobrovnikov, A.R.Buzykaev, G.M.Kolachev,\\
S.A.Kononov, E.A.Kravchenko, G.D.Minakov, A.P.Onuchin\thanks{onuchin@inp.nsk.su},\\
Budker Institute of Nuclear Physics, Novosibirsk, Russia\\
A.F.Danilyuk,\\
Boreskov Institute of Catalysis, Novosibirsk, Russia\\
F.F.Guber, A.B.Kurepin,\\
Institute of Nuclear Research, Moscow, Russia\\
V.A.Krasnov,\\
Joint Institute of Nuclear Research, Dubna, Russia
}

\maketitle

\newcommand{\piK}{$\pi$/K}
\begin{abstract}
The work is devoted to the development of the aerogel Cherenkov
counters with the light collection on wavelength shifters and PMTs (ASHIPH).
The ASHIPH system has been developed for the KEDR detector.
Tests of the counters have been carried out on the Dubna accelerator,
the \piK\ separation obtained is about 4$\sigma$.

The ASHIPH system is suggested for the PEP-N detector.
\end{abstract}

\section{Introduction}
The collaboration of Budker Institute of Nuclear Physics and Boreskov
Institute of Catalysis has started the development of aerogel Cherenkov
counters in 1986 \cite{INSTR90} at Novosibirsk. The most important results
of this work
are the development of ASHIPH counters, development of the Monte Carlo code
for simulation of Cherenkov light collection,
and production of aerogel with high optical parameters.

The idea of the ASHIPH method is to use light guides with the
wavelength shifting admixture for the light collection on PMT \cite{ASHIPH}.
As compared with the direct light collection on PMT, the ASHIPH
method allows the number of PMTs to be reduced essentially \cite{INSTR96,mont98,vienna98,INSTR99}.

In order to simulate the processes of the light collection and propagation
inside the aerogel Cherenkov counter a special code was developed in
Budker Institute of Nuclear Physics \cite{INSTR90,ASHIPH,chep97}. This
code simulates the following processes: Rayleigh scattering inside the
aerogel, Lambert angular distribution of the reflected light from the
walls, Fresnel refraction on the boundary of two continuous media, and
a light absorption inside the aerogel and on the walls.
Using this code we optimize the counter design and calculate the number of
photoelectrons without production of series of prototypes.

\begin{figure}[htb]
\centering
\includegraphics*[width=60mm]{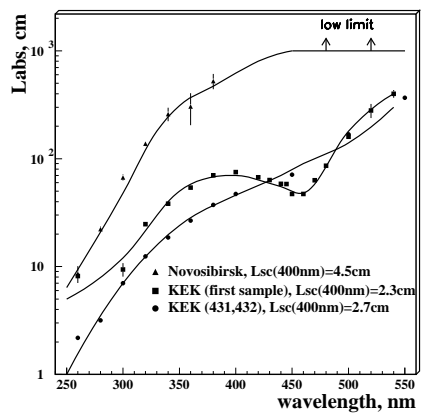}
\caption{Labs and Lsc (in comments) for Novosibirsk and KEK aerogels.}
\label{fig:Labs}
\end{figure}

We use the high optical properties aerogel SAN-96
\cite{Danilyuk,RICH98,Synthesis} produced at Novosibirsk by
collaboration of Boreskov Institute of Catalysis and Budker Institute
of Nuclear Physics. The data on the light absorption (Labs) and
scattering (Lsc) lengths in the SAN-96 aerogel are shown in
Figure~\ref{fig:Labs}.
The data for aerogel samples produced at KEK(Japan) \cite{KEK}
are also presented \cite{INSTR99}.

\section{ASHIPH for KEDR}
\subsection{Layout}

\begin{figure}[htb]
\centering
\includegraphics*[width=60mm]{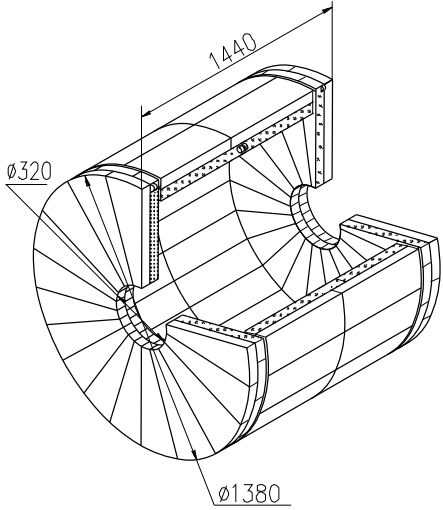}
\caption{The KEDR ASHIPH system.}
\label{fig:ATC}
\end{figure}

\begin{figure}[htb]
\centering
\includegraphics*[width=65mm]{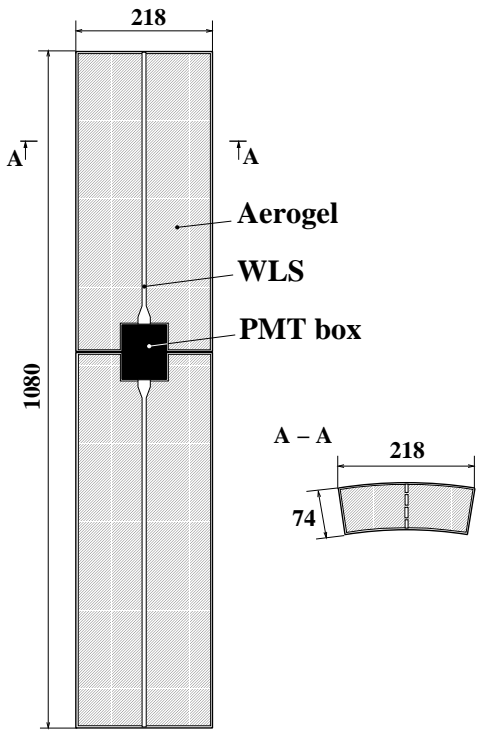}
\caption{The barrel ASHIPH counter.}
\label{fig:bar}
\end{figure}

\begin{figure}[htb]
\centering
\includegraphics*[width=60mm]{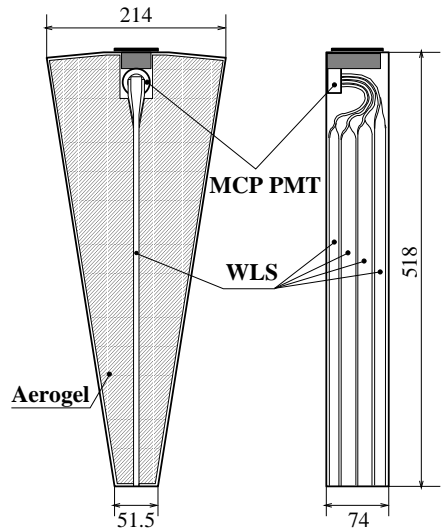}
\caption{The endcap ASHIPH counter.}
\label{fig:ecap}
\end{figure}

The ASHIPH system \cite{INSTR99} for the KEDR detector \cite{KEDR2001}
is shown in Figure~\ref{fig:ATC}. The system comprises 160
counters: 80 barrel (Fig.~\ref{fig:bar}) and 80 endcap
counters (Fig.~\ref{fig:ecap}). The total volume of aerogel
is 800 liters. The use of aerogel with the
refractive index 1.05 gives the possibility to separate pions and kaons
in the momentum range $0.6\div 1.5$~GeV/c.

An important feature of the project is a two-layer design. The counter
are arranged in such a way that a particle from the interaction point
with a momentum above 0.6 GeV/c does not cross shifters in both
layers simultaneously. It is possible to use the information
from both layers for the essential part of the particles.

\subsection{Reflector}
\begin{figure}[htb]
\centering
\includegraphics*[width=60mm]{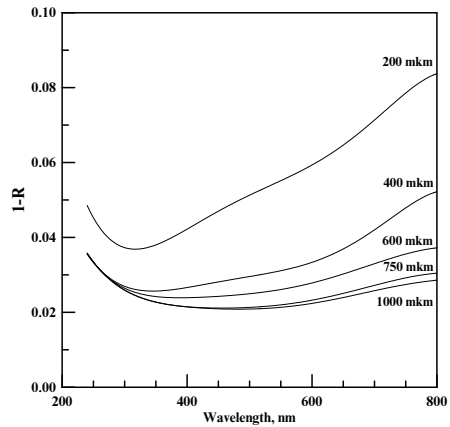}
\caption{Reflection R of PTFE teflon.}
\label{fig:PTFE}
\end{figure}

We use the multi-layer PTFE film from Tetratex company as a
reflector. The results of measurement \cite{INSTR96}
of the reflection coefficient for the different thickness of
reflector are shown in Figure~\ref{fig:PTFE}. The PTFE teflon has
four times larger radiation length than the KODAK paint.
Due to using teflon the amount of material in front of
the calorimeter is significantly decreased.

\subsection{BBQ wavelength shifter}
\begin{figure}[htb]
\centering
\includegraphics*[width=60mm]{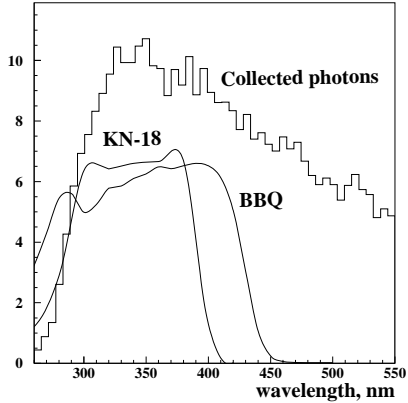}
\caption{The absorption spectra of BBQ and KN-18 (POPOP) wavelength shifters,
collected Cherenkov photons spectrum.}
\label{fig:BBQabs}
\end{figure}

The absorption spectrum of BBQ is presented in
Figure~\ref{fig:BBQabs} together with the spectrum of collected
Cherenkov photons and the absorption spectrum of POPOP (KN-18).
The production of plexiglass plates doped with BBQ
was mastered in Institute of Polymers at Dzerzhinsk. The cutting,
polishing, and twisting were organized in Budker Institute of Nuclear
Physics \cite{INSTR99}.

\subsection{Microchannel plate PMT}
\begin{figure}[htb]
\begin{center}
\includegraphics[width=60mm]{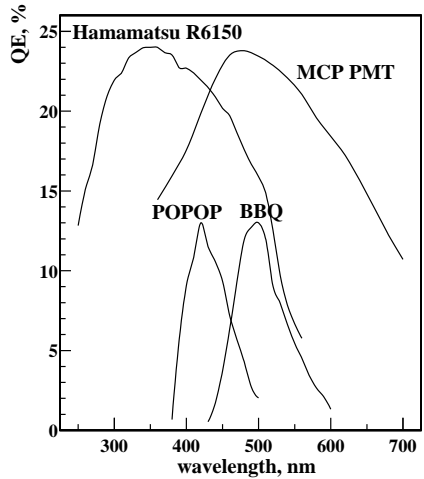}
\caption{Hamamatsu R6150 $N^{\underline{\circ}}$ZH2673 and Katod MCP PMT
$N^{\underline{\circ}}$1570 quantum efficiencies. POPOP and BBQ emission spectra.}
\label{fig:qe}
\end{center}
\end{figure}

We use microchannel plate (MCP) PMTs with multialkali photocathode
produced in Novosibirsk by ``Ekran'' plant \cite{vienna98,INSTR99}.
The size of this device is small: 31 mm diameter and 17 mm thickness.
The photocathode is 18 mm in diameter.
Our measurements of quantum efficiency (QE) of this PMT
and the FM PMT R6150 are shown in Figure~\ref{fig:qe}.
The shift of spectral response to the region of longer wavelengths
is the essential advantage of MCP PMTs in respect of detecting BBQ emission.

The decrease of
MCP PMT multiplication gain in high magnetic field is not so strong as
for fine mesh PMTs of Hamamatsu. In the magnetic field of 1.5~Tesla
the gain drops in some 5 times.
\subsection{Amount of material in the system}

The amount of material in the KEDR ASHIPH system is shown in
Table~\ref{tab:amofmat}.

\begin{table}[htb]
\begin{center}
\caption{The amount of material in the ACC system.}
\begin{tabular}{|l|c|c|c|}   \hline
        & Density, g/cm$^3$    	& X$_0$, cm	& \% of X$_0$\\ \hline
Aerogel & 0.243                 & 112           & 6     \\
Frame (Al) & 2.7                & 8.9           & 4.3     \\
WLS 	& 1.49                 	& 34.4          & 0.3     \\
PTFE 	& 2.2                 	& 15.8          & 1.1     \\
MCP PMT &                  	&               & 0.2     \\ \hline
Total 1 layer&                  &            	& 11.9     \\ \hline
Total 2 layers&                 &            	& 24     \\ \hline
\end{tabular}
\label{tab:amofmat}
\end{center}
\end{table}

\section{Test beam results for the KEDR ASHIPH}

\subsection{Layout of the experiment}
The endcap counters of the KEDR detector were tested on the particle
beam at the Dubna accelerator. The counters were filled with blocks
of SAN-96 aerogel.

The tested counters were turned relative to the beam axis,
so that particles traverses the counter at the same angle as in the
KEDR detector.
The time-of-flight counters with 30 m base were used to separate beam particles.
A hodoscope of scintillation counters was used to determine coordinates,
with which a particle passes the counter.
The measurements were carried out with protons of
0.86 GeV/c--2.1~GeV/c momentum and pions of
0.86 GeV/c--1.6~GeV/c momentum.

\subsection{Number of photoelectrons, counter homogeneity}
\begin{figure}[hb]
\centering
\includegraphics*[width=70mm]{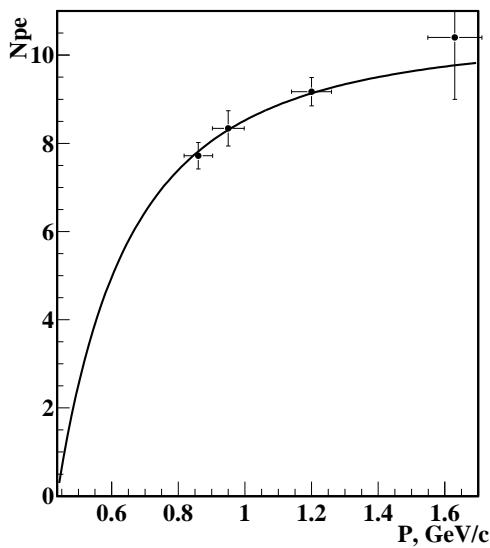}
\caption{The number of photoelectrons for pions versus momentum.
The curve represents the theoretical formula fit.}
\label{fig:Npe}
\end{figure}

The dependence of the detected Cherenkov light on momentum was measured with
pions (Fig.~\ref{fig:Npe}).
The theoretical fit to the experimental point gives the resulting number
of photoelectrons at $\beta=1$ as 10.6 pe.

\begin{figure*}[ht]
\centering
\vspace{-1.5cm}
\includegraphics*[width=140mm]{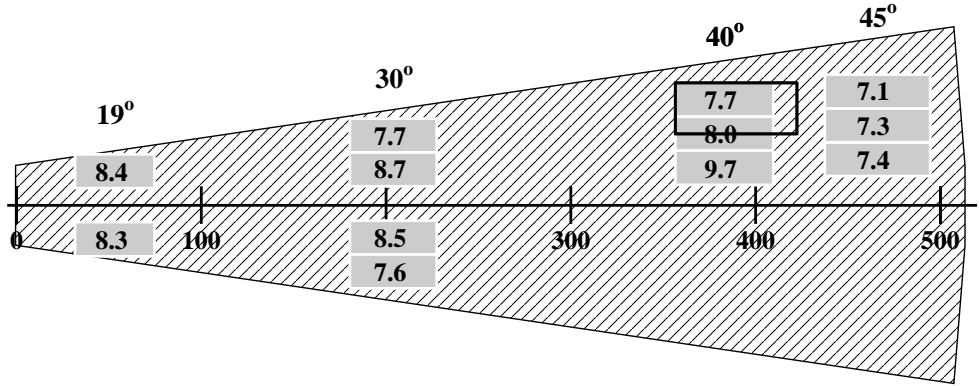}\\[-2cm]
\caption{The number of photoelectrons in different points of the counter.
The open rectangle designates the area where the \piK\ separation was measured.}
\label{fig:unif}
\end{figure*}

The homogeneity of the light collection was measured with the
0.83~GeV/c pions over the whole area of the counter.
As shown in the Figure~\ref{fig:unif} the signal varies from 7.1 pe
up to 9.7 pe.

\subsection{\piK\ separation}
\begin{figure}[htb]
\centering
\includegraphics*[width=70mm]{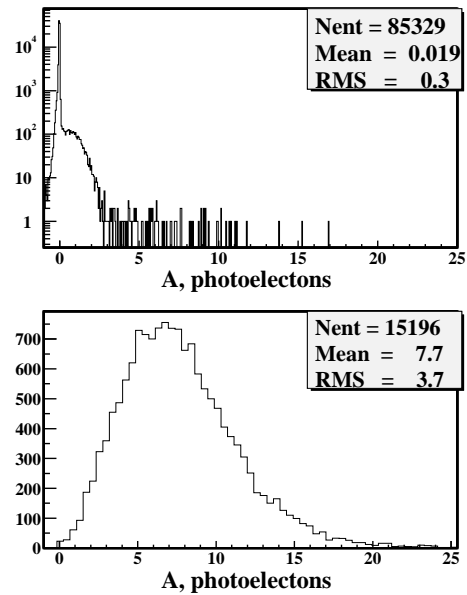}
\caption{Amplitude spectra for kaons (top) and pions (bottom), P=0.86~GeV/c.}
\label{fig:ampsp}
\end{figure}

The data for kaons were obtained using protons with
corresponding velocity.
Figure~\ref{fig:ampsp} illustrates ``kaon''
and pion amplitude spectra obtained from the counter at P=0.86~GeV/c.

\begin{figure}[htb]
\centering
\includegraphics*[width=70mm]{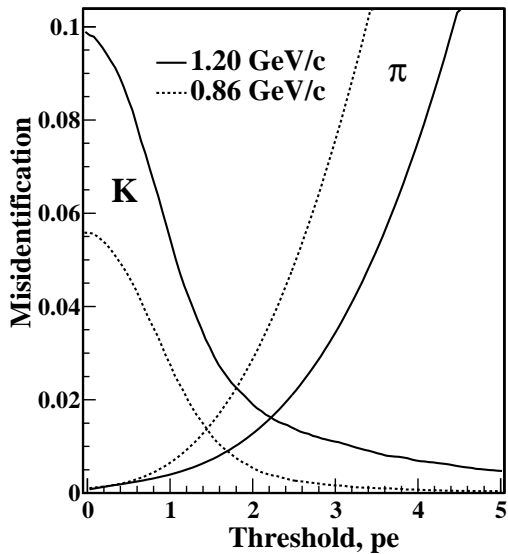}
\caption{Probability of misidentification for
kaons and pions as a function of threshold.
P=0.86~GeV/c, 1.2~GeV/c.}
\label{fig:effcnt1}
\end{figure}

In Figure~\ref{fig:effcnt1}
the probabilities of kaon and pion misidentification are shown as a
function of threshold for 0.86 and 1.2 GeV/c momenta.
For 0.86 GeV/c at the zero threshold on the amplitude the signal
pion suppression factor is 860, with kaon detection efficiency being equal to
94\% (separation is 4.7 $\sigma$). And for 1.2 GeV/c pion suppression factor
is 1300, with kaon detection efficiency being 90\%\ (4.5 $\sigma$).

We would like to note that the aerogel counters system of the KEDR detector
includes two layers and most of the particles will cross two counters
in good conditions. For these particles the identification will be better.

\section{ASHIPH for PEP-N}

\begin{figure*}[t]
\centering
\includegraphics*[width=140mm]{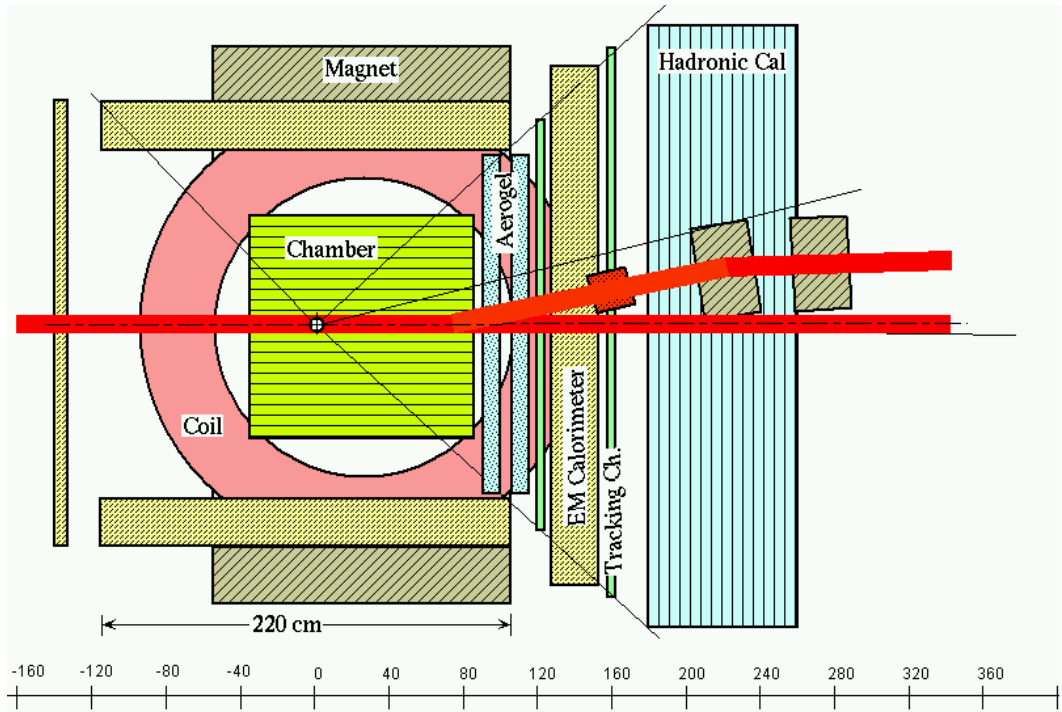}
\caption{Top view of the PEP-N detector.}
\label{fig:topview}
\end{figure*}

\begin{figure*}[t]
\centering
\includegraphics*[width=140mm]{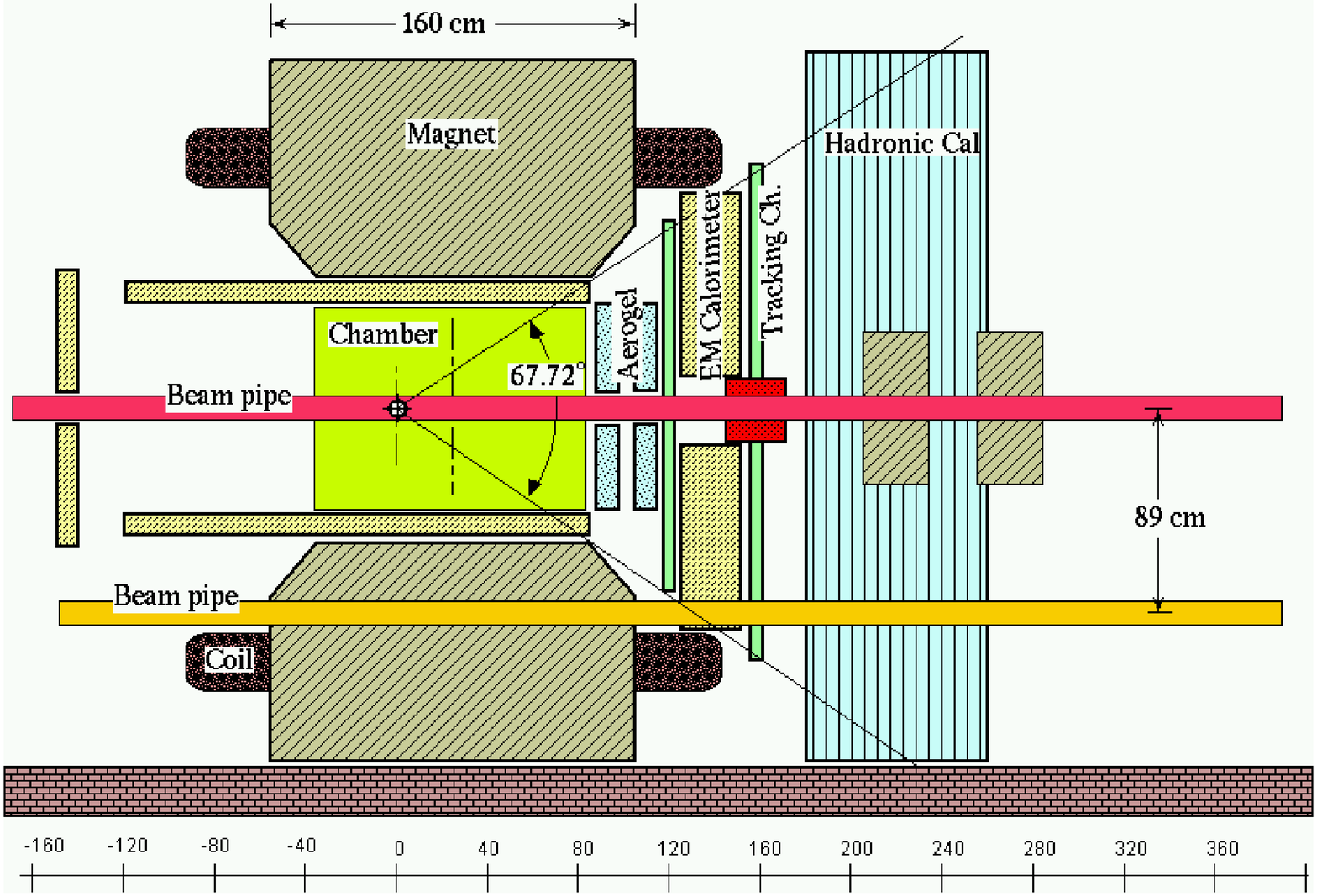}
\caption{Side view of the PEP-N detector.}
\label{fig:sideview}
\end{figure*}

The PEP-N experiment is proposed for the investigation of $e^+e^-$
collisions in the c.m. energy from 1.4 to 3~GeV. The Aerogel Cherenkov
Counter system is provided for \piK\ separation in the forward direction
of the detector (Figs.~\ref{fig:topview},\ref{fig:sideview}).

The proposed system is analogous to the KEDR ASHIPH system.
Total volume of aerogel is 350 liters. We suggest to use MCP PMTs in
the counters. The refractive index of aerogel is 1.05. This provides the
\piK\ identification in the momentum range from 0.6 to 1.5~GeV/c.
Identification below 0.6~GeV/c is provided by dE/dX
in coordinate system and TOF measurements with calorimeter.

\begin{figure}[htb]
\centering
\includegraphics*[width=60mm]{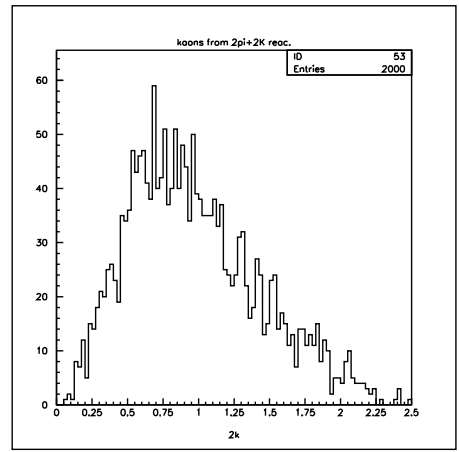}
\caption{The momentum distribution of kaons in
laboratory coordinates in the reaction $e^+e^- \rightarrow K^+K^-\pi^+\pi^-$
at E$_{\mathrm{c.m.}}$~=~2~GeV.}
\label{fig:plab_K}
\end{figure}

The Monte Carlo calculations were performed for the $e^+e^- \rightarrow
K^+K^-\pi^+\pi^-$ reaction at E$_{\mathrm{c.m.}}$~=~2~GeV.
The momentum distribution of kaons in
laboratory coordinates is shown in Figure~\ref{fig:plab_K}.
One can see that the identification region of the ASHIPH system covers
the main part of the events.

The results on identification acceptance for 4 track events is presented
in Table~\ref{tab:IDaccept}. The tracker acceptance for this reaction is
88\% for 4 tracks and 12\% for 3 tracks.

\begin{table}[htb]
\begin{center}
\caption{Identification acceptance for 4 track events. N$_{\mathrm{ID}}$ is the number of particles identified.}
\begin{tabular}{|l|l|l|} \hline
N$_{\mathrm{ID}}$ & Identification systems & Acceptance \\ \hline
4 	& dE/dX + TOF 		& 2\% 	\\
4 	& dE/dX + TOF + ASHIPH	& 48\%	\\
$\ge$3 	& dE/dX + TOF 		& 41\% 	\\
$\ge$3 	& dE/dX + TOF + ASHIPH	& 93\%	\\ \hline
\end{tabular}
\label{tab:IDaccept}
\end{center}
\end{table}

The amount of material in the PEP-N ASHIPH system is about 20\% of $X_0$.

\section{Conclusion}
The use of high transparency aerogel, ASHIPH method together with the
detailed Monte Carlo calculations have helped us to develop and construct
the aerogel Cherenkov counters for the KEDR detector. The counters have been
tested on beam, 4$\sigma$ \piK\ separation has been obtained.

The Aerogel Cherenkov Counters system is proposed for the PEP-N
detector. This system is analogous to one developed for the
KEDR detector. It provides 4$\sigma$ \piK\ separation in
the momentum range $0.6\div 1.5$~GeV/c. Total volume of aerogel is
350 liters, amount of material in front of the calorimeter is
about 20\% of $X_0$.

\end{document}